\documentclass[aps,prb,twocolumn,showpacs]{revtex4-1}
\usepackage{graphics}
\usepackage{amsmath}
\usepackage{amsfonts}
\usepackage{bm}
\usepackage{color}
\usepackage{bbm}

\newcommand{\be}{\begin{equation}}
\newcommand{\ee}{\end{equation}}
\def\boldsymbol#1{\mbox{\boldmath$#1$}}
\newcommand{\q}{\mathbf{q}}
\newcommand{\p}{\mathbf{p}}

\newcommand{\n}{\mathbf{n}}
\renewcommand{\r}{\mathbf{r}}

\newcommand{\beq}{\begin{equation}}
\newcommand{\eeq}{\end{equation}}
\newcommand{\beqa}{\begin{eqnarray}}
\newcommand{\eeqa}{\end{eqnarray}}
\newcommand{\bea}{\begin{eqnarray}}
\newcommand{\eea}{\end{eqnarray}}
\usepackage{amssymb}
\usepackage{array}
\usepackage{amsmath}
\usepackage{graphicx}\usepackage{graphics}
\usepackage{dcolumn}
\usepackage{bm}\usepackage{varioref}

\begin{document}

\title{Magneto-transport phenomena  related to the chiral anomaly in Weyl semimetals}

\author{B. Z. Spivak}

\affiliation{Department of Physics, University of Washington, Seattle, WA 98195 USA}

\author{A. V. Andreev}

\affiliation{Department of Physics, University of Washington, Seattle, WA 98195 USA}

\date{\today}

\begin{abstract}
We present a theory of magnetotransport phenomena related to the
chiral anomaly in Weyl semimetals.
We show that conductivity, thermal conductivity, thermoelectric and the sound absorption coefficients exhibit
strong and anisotropic magnetic field dependencies.   We also discuss properties of magneto-plasmons and magneto-polaritons, whose existence is  entirely determined by the chiral anomaly. Finally, we discuss the conditions of applicability of the quasi-classical description of electron transport phenomena related to the chiral anomaly.
\end{abstract}


\maketitle

\addtolength{\abovedisplayskip}{-1mm}

\noindent

\section{Introduction}

A concept of zero band gap semiconductors with topologically protected massless Dirac
points (Weyl semi-metals) has been introduced in Refs.~\onlinecite{TurnerVishwanath,BurkovBalents,HalaszBalents,BurkovHookBalents} (see also for a review Ref.~\onlinecite{VafekVishwanath}).
In these materials the valence and conduction bands touch at isolated points in the Brilluoin zone, and the electron states near these points may be described by  the Dirac Hamiltonian,
\begin{equation}\label{eq:H_Dirac}
  H= k^{(a)}v\, \p \cdot \boldsymbol{\sigma},
\end{equation}
where $\sigma_i$ are Pauli matrices. The coefficient $ k^{(a)}=\pm 1$ in Eq.~(\ref{eq:H_Dirac}) indicates the
handedness or chirality of each Weyl node, labeled by $a$.
Due to Nielsen-Ninomiya theorem \cite{NielsonNyninomiya} the number of these Weyl points $\mathcal{N}_{v}$ should be even, and numbers of opposite chirality nodes should be equal.  The stability of Weyl nodes is related to the fact that the flux of Berry curvature through a closed surface  surrounding the node is quantized.
Since the time reversal symmetry requires  the Berry curvature to be an odd function of momentum and inversion symmetry requires  it to be even,  Weyl nodes can only exist in crystals with either broken inversion or time reversal symmetry. In the former case the minimal number of Weyl nodes is four, while in the latter case it is two.
An interesting feature of the system with the massless Dirac electron spectrum is the existence of the chiral anomaly.~\cite{Adler,Jackiw} One of its manifestation is  a giant and strongly anisotropic negative magneto-resistance which exists in the case when the electric and the magnetic fields are  collinear.
  It was predicted by Nielsen and Ninomiya \cite{NielsonNyninomiya} in the ultra-quantum  regime of strong magnetic field, where only zeroth Landau level is partially occupied. This phenomenon is related to the fact that in the presence of a magnetic field the electrons can be transferred between different Weyl nodes by the spectral flow caused by the electric field parallel to the magnetic one.  It was shown in Ref.~\onlinecite{SonSpivak} that  the strongly anisotropic magnetoresistance  due to the chiral anomaly persists to the semiclassical regime of weak magnetic fields, where it can be described by the Boltzmann kinetic equation. These effects are related to the ``chiral magnetic effect'' which may be observable in relativistic heavy-ion collisions~\cite{Karzeev}.

Recently in a series of remarkable experiments~\cite{Qiang Li_Kharzeev,JunXiong_Ong,Jun Xiong,Hasan,Xiaojun Yang,ChenZangEnzeZhangYanwenLiu,Shekhar} a significant and strongly anisotropic magnetoresistance has been measured in a number of systems. It is  negative for the case where the magnetic and electric fields are parallel and positive in the case when the electric and magnetic field are perpendicular. This
 can be interpreted as a strong evidence of existence of the chiral anomaly in these materials.

In this article we consider electron transport phenomena related to the chiral anomaly in several physical regimes. We predict a
strong and anisotropic magnetic field dependencies of thermal conductivity and the thermoelectric the sound absorption coefficients. We also discuss  new types of magneto-plasmons and magneto-polaritons in Weyl metals.

\section{Description of electron transport phenomena related to the chiral anomaly \label{sec:formalism}}

We assume that the coupling constant for the electron-electron interaction is small, $\alpha=e^{2}/\varepsilon\hbar v\ll 1$, where $\varepsilon$ is the dielectric constant.
 In this case one can neglect the renormalization of the electron spectrum
and  develop a  scheme to describe
transport phenomena in Weyl semi-metals which is based on the Boltzmann kinetic equation.

In the presence of an external  magnetic field the momentum operator $\p$ in Eq.~(\ref{eq:H_Dirac})  becomes $\p=- i \hbar \boldsymbol{\nabla} - \frac{e}{c}\mathbf{A}$, where $\mathbf{A}$ is the vector potential.
For a uniform magnetic field $\mathbf{B}=(0,0,B)$ the  spectrum of the massless Dirac Hamiltonian Eq.~(\ref{eq:H_Dirac})  is well known,
\begin{equation}
\epsilon_{n}(p_{z})=\left\{ \begin{array}{cc}
                              \pm v\sqrt{2n\frac{\hbar e}{c}B+p_{z}^{2}}, &  n=1, 2,\ldots,\\
                            k^{(a)} v p_z, & n=0.
                            \end{array}
\right.
\end{equation}
Here $p_{z}$ is the momentum in the direction of the magnetic field.
A peculiarity of this spectrum is that in addition to $n\neq 0$ Landau levels
there is a ``chiral", $n=0$, Landau level with an asymmetric in $p_z$ dispersion.

If the electron level broadening $\gamma$ is smaller than the Landau level spacing one can describe the electron transport phenomena with the aid of the Boltzmann kinetic equation for the electron distribution function of the quantized Landau orbitals. At low magnetic fields and large characteristic electron energies this requirements is violated. In this regime the Landau quantization may be neglected and electron dynamics  may be described using the quasiclassical approach. The quasi-classical equations of electron motion were generalized in Ref.~\onlinecite{SundaramNiu} to include effects of Berry curvature that arises in crystals with broken inversion or time reversal symmetry,
\begin{subequations}
\label{eq:EqMotion}
\begin{eqnarray}\label{eq:r_dot}
{\bf \dot{r}}&=&\frac{\partial \epsilon_{{\bf p}}}{\partial {\bf p}}+{\bf \dot{p}} \times {\bf \Omega}^{(a)}_{{\bf p}},  \\
\label{eq:p_dot}
{\bf \dot{p}}&=&e{\bf E}-\Lambda_{ij}^{(a)}({\bf p})\nabla_{\bf r}u_{ij}({\bf r})+\frac{e}{c}\, {\bf \dot{r}}\times {\bf B}.
\end{eqnarray}
\end{subequations}
Here the index $a$ labels the valleys, ${\bf \Omega}^{(a)}_{{\bf p}}=\nabla_{{\bf p}} \times {\bf A}^{(a)}_{{\bf p}}$ is the Berry curvature,  ${\bf A}^{(a)}_{{\bf p}}= i\langle u^{(a)}_{\bf p}|\nabla_{\bf p}| u^{(a)}_{\bf p}\rangle $,  $|u^{(a)}_{{\bf p}}\rangle $ are the Bloch wave functions, $\Lambda^{(a)}_{ij}({\bf p})$ is the deformation potential in valley $a$,  $u_{ij}({\bf r},t)=\frac{1}{2}(\partial_{i}u_{j}({\bf r},t)+\partial_{j}u_{i}({\bf r},t))$ is the strain tensor,  with $u_{i}({\bf r},t)$ being the components of the lattice displacement, and $c$ the speed of light.
The second term in Eq.~(\ref{eq:p_dot}) for ${\bf \dot{p}}$ represents the  force
 exerted on the electrons by the strain, $-\nabla \delta \epsilon^{(a)}_{{\bf p}}=- \Lambda^{(a)}_{ij}({\bf p}) \boldsymbol{\nabla}_\r u_{ij}({\bf r},t)$.

The quasiclassical Boltzmann kinetic equation for the electron distribution function $n_{{\bf p}}({\bf r},t)$ that follows from Eq.~(\ref{eq:EqMotion}) has the following form,
\begin{equation}\label{KinEq}
 \frac{\partial n^{(a)}_{{\bf p}}}{\partial t}+{\bf \dot{r}}\cdot \frac{\partial n^{(a)}_{{\bf p}}}{\partial {\bf r}}+
{\bf \dot{p}}\cdot \frac{\partial n^{(a)}_{{\bf p}}}{\partial {\bf p}}=I^{(a)}_{st}\{n^{(a)}_{{\bf p}}\},
\end{equation}
while the expression for the particle current density in an individual valley is given by
\begin{eqnarray}
 {\bf j}^{(a)}\!\!&=&\!\int \frac{d^3 {\bf p}}{(2\pi \hbar)^{3}}\left(1+\frac{e}{c}{\bf \Omega}^{(a)}_{{\bf p}} \cdot{\bf B}\right) {\bf \dot{r}} \, n^{(a)}_{{\bf p}} \nonumber \\
&\!=\!&\!\!\int\!\! \frac{d^3 {\bf p} \,  n^{(a)}_\p }{(2\pi \hbar)^{3}}\left[\frac{\partial \epsilon_{{\bf p}}}{\partial  {\bf p}}\!+\!e{\bf E}\!\times\! {\bf \Omega^{(a)}_{{\bf p}}}\!+\!\frac{e {\bf B}}{c}\!\left(\!\Omega^{(a)}_{{\bf p}}\!\cdot \!\frac{\partial \epsilon_{{\bf p}}}{\partial {\bf p}}\!\right)\!\right] \label{eq:current}.
\end{eqnarray}
The second line in Eq.~(\ref{eq:current}) is obtained by substituting the solution of the classical equations of motion (\ref{eq:EqMotion}) into the first line.
The collision integral $I^{(a)}_{st}\{n^{(a)}_{{\bf p}}\}$ in Eq.~(\ref{KinEq}) accounts for the evolution of the distribution function due to various electron scattering processes.

It is important to distinguish between inter-valley scattering processes, which transfer electrons between the valleys, and intra-valley scattering which does not change the number of electrons in a given valley.
I the regime where the elastic  intra-valley relaxation time $\tau_{intra}$ is the shortest relaxation time in the problem, the description  of electron transport can be considerably simplified. In this case one can neglect the intravalley anisotropy of the electron distribution  and describe the system by the distribution function  $n^{(a)}_\p=n^{(a)}(\epsilon_\p)$, which depends only on the electron energy and the valley index  $a$.
 In the remainder of the paper we specialize to this regime.
Furthermore, we assume that inter-valley scattering is dominated by elastic scattering of electrons from impurities,
 which allows us to write the corresponding part of the scattering integral in the relaxation time approximation with
 the energy-dependent inter-valley scattering relaxation time $\tau (\epsilon)$. Then the Boltzmann equation (\ref{KinEq}) simplifies to
\begin{widetext}
\begin{equation}\label{epsilonKinEq}
\frac{\partial n^{(a)}(\epsilon)}{\partial t}+\frac{1}{\nu^{(a)}(\epsilon)}\boldsymbol{\nabla}\cdot {\bf j}^{(a)}(\epsilon)=-\frac{k^{(a)}}{\nu^{(a)}(\epsilon)}\frac{e}{4\pi^{2}\hbar^{2}c}\left( e{\bf E} -
\bar{\Lambda}_{ij}^{(a)}\nabla_{\bf r}u_{ij}({\bf r})\right)\cdot {\bf B}\, \frac{\partial n^{(a)}(\epsilon)}{\partial \epsilon}-
\frac{n^{(a)}(\epsilon)-\bar{n}(\epsilon)}{\tau(\epsilon)}+I^{(a)}_{\epsilon}.
\end{equation}
\end{widetext}
Here $k^{(a)}$ denotes the  quantized flux of the Berry curvature through the constant energy surface,
\begin{equation}\label{eq:Berry_flux}
 k^{(a)}=\frac{1}{2\pi \hbar }\int {\bf \Omega}^{(i)}_{\bf p} \cdot d {\bf S}^{(a)}= 0, \pm 1, \ldots,
\end{equation}
where $ d\mathbf{S}^{(a)}$ is the area differential. For Weyl semimetals the Berry curvature flux is given by the valley helicity  $k^{(a)}=\pm 1$ in Eq.~(\ref{eq:H_Dirac}).  The quantity $\bar{\Lambda}_{ij}^{(a)}(\epsilon)$ in Eq.~(\ref{epsilonKinEq})
 is the deformation potential  averaged over the direction of momentum ${\bf p}$  at a given energy $\epsilon$ in the $a$-th valley, $\bar{n}(\epsilon)=\sum_{a}n^{(a)}(\epsilon)/\mathcal{N}_v$ (with $\mathcal{N}_v$ being the number of valleys) is the  electron distribution function averaged over the valleys.
The collision integral $I^{(a)}_\epsilon$ in Eq.~(\ref{epsilonKinEq}) describes the inelastic scattering processes.
The expression for the density of states in the $a$-th valley has the form
\begin{equation}\label{eq:nu_def}
\nu^{(a)}(\epsilon)=\int \frac{d^3 {\bf p}}{(2\pi \hbar)^{3}} \left(1+\frac{e}{c}{\bf \Omega}^{(a)}_{{\bf p}}\cdot{\bf B}\right)\delta(\epsilon-\epsilon_{{\bf p}})\sim \frac{\epsilon^{2}}{v^{3}2\pi^{2}\hbar^{3}},
\end{equation}
where  we can neglect the weak magnetic field dependence. The particle flux density at a given energy  in valley $a$ in Eq.~(\ref{eq:current}) can be expressed as
\begin{eqnarray}
 \label{epsilonCurrent}
  {\bf j}^{(a)} ({\epsilon})&= & \! n(\epsilon)\!\!\int\! \frac{d^3 \p   \delta (\epsilon_\p -\epsilon)}{(2\pi \hbar)^3} \left[\!e{\bf E}\!\times\! {\bf \Omega^{(a)}_{{\bf p}}}\!+\!\frac{e {\bf B}}{c}\!\left(\!\Omega^{(a)}_{{\bf p}}\!\cdot \!\frac{\partial \epsilon_{{\bf p}}}{\partial {\bf p}}\!\right)\!\right] \nonumber  \\
  &=& k^{(a)}\frac{e}{4\pi^{2}\hbar^{2}c} \, {\bf B}\, n^{(a)}(\epsilon).
  \end{eqnarray}
The densities of electric current $\mathbf{j}$  and thermal flux ${\bf I}^{(a)}_{\epsilon}$ may be expressed in terms of $ {\bf j}^{(a)} ({\epsilon})$  as
\begin{eqnarray}\label{IntegralCurrent}
{\bf j}&=&\sum_a e\int {\bf j}^{(a)} ({\epsilon}) d\epsilon,\\
\label{IntegralCurrentEnergy}
{\bf I}_{\epsilon}&=& \sum_a \int (\epsilon-\mu) {\bf j}^{(a)} ({\epsilon})d\epsilon.
\end{eqnarray}
In conventional conductors the assumption of complete intra-valley momentum relaxation ($\tau_{intra}\rightarrow 0$) would result in a vanishing contribution of each valley to the current density (\ref{IntegralCurrent}) and thermal flux density (\ref{IntegralCurrentEnergy}).
In Weyl metals the particle flux  ${\bf j}^{(a)}({\epsilon})$ in a given valley does not vanish even in this limit.\cite{Vilenkin,Karzeev}
The nonvanishing particle flux density  (\ref{epsilonCurrent}) at full momentum relaxation is a manifestation of the chiral anomaly in the semiclassical regime.

We note that the expression for the contribution of a given valley into the  current density in  Eq.~(\ref{IntegralCurrent}) is ambiguous.  It depends on the choice of the lower cutoff of the energy integral. The observable net current however is given by the sum of valley currents and is therefore independent of this choice.

We note that the first term  in the right hand side of Eq.~(\ref{epsilonKinEq}) describes the evolution of the electron distribution due to spectral flow, whereby electrons in the chiral, $n=0$, Landau level
are accelerated by the electric field. The presence  of the density of states
in the denominator of this term is related to the fact that the flux of electrons due to the spectral flow is scattered by intra-valley processes into all available states at a given energy $\epsilon$.

The system of Eqs.~(\ref{epsilonKinEq}), (\ref{epsilonCurrent}), (\ref{IntegralCurrent}), (\ref{IntegralCurrentEnergy}), can describe the electron transport not only in the semiclassical regime but also at arbitrary magnetic fields. In  the latter case the density of states $\nu^{(a)}(\epsilon)$ should account for Landau quantization. For example,  in  the  ultra-quantum limit where only the chiral ($n=0$) Landau leven matters,
 $\nu^{(a)}(\epsilon)=e B/( 4 \pi^2 \hbar^2 cv)$. The only restriction for validity of the system is that the characteristic electron energies should be larger than the broadening of electron levels. We will see below that sometimes this requirement may be violated even at relatively high temperatures.
 Below we apply this kinetic scheme to study several transport phenomena in Weyl semimetals.

\section{magnetoresistance and magneto-thermal conductivity\label{sec:transport}}

We begin by considering the longitudinal conductivity, which corresponds to the situation  where the external electric field is directed along the magnetic field. In the ultra-quantum limit, $v/L_{B}\gg T, \mu, \gamma$, and in the single particle approximation the longitudinal conductivity of Weyl metals was obtained in Ref.~\onlinecite{NielsonNyninomiya},
\begin{equation}\label{sigmaUltraQuant}
\sigma_{zz}=\mathcal{N}_{v}\frac{e^{2}v}{4\pi \hbar L_{B}^{2}}\tau(\epsilon=0,B).
\end{equation}
Here $L_{B}=\sqrt{\hbar c/eB}$ is the magnetic length.
All other components of the conductivity tensor in this approximation are zero.
The magnetic field dependence of the  longitudinal magnetoresistivity in this regime is controlled by the corresponding dependence of the inter-valley scattering rate $1/\tau(\epsilon =0, B )$. The latter  is non-universal and depends on the type of impurities. Its evaluation is not essentially different from the calculation of the backscattering time in conventional semiconductors in the ultra-quantum limit (see e.g. Refs.~\onlinecite{ArgyresAdams,Murzin}).
For example, for the case of short-range impurities the intervalley scattering rate is proportional to the electron density of states $1/\tau \sim n_i U_0^2/(v L_B^2)$, where  $U_0$ and $n_i$ are respectively the strength and density of impurities.
Thus in this case the conductivity becomes independent of the magnetic field.~\footnote{Note that in conventional semiconductors with short range impurities the longitudinal magnetoresistance is positive in the ultra-quantum limit.~\cite{Murzin,Sarma}.}

Let us now consider the opposite, quasiclassical regime, in which the electrical conductivity is dominated by electrons with large energies.
Neglecting inelastic scattering, linearizing Eq.~(\ref{epsilonKinEq}), and using Eqs.~(\ref{epsilonCurrent}) and (\ref{IntegralCurrent}) at $u_{ij}=0$ we get the following expression for the longitudinal conductivity,
\begin{equation}\label{sigmaFormQuasi}
\sigma_{zz}=\mathcal{N}_{v}\left(\frac{e^{2}}{4\pi^{2}\hbar^{2}c}\right)^2 B^{2}\int d \epsilon \frac{\tau(\epsilon)}{\nu(\epsilon)}\left( - \frac{\partial n_{F}(\epsilon)}{\partial \epsilon}\right),
\end{equation}
where $n_{F}(\epsilon)=1/(e^{(\epsilon-\mu)/T}+1)$ is the equilibrium Fermi distribution function.

\emph{Quantum degenerate regime.}  At $\mu\gg T$  the  expression (\ref{sigmaFormQuasi}) for the longitudinal conductivity  reduces to~\cite{SonSpivak}
\begin{equation}\label{sigmaLargeMu}
\sigma_{zz}(\mu)=\mathcal{N}_{v}\frac{e^{2}}{8\pi^{2}\hbar c}\frac{(eB)^{2}v^{2}}{\mu^{2}}\frac{v}{c}\tau(\mu).
\end{equation}
Though Eq.~(\ref{sigmaFormQuasi}) was obtained under the assumption of absence of inelastic scattering, the result for the conductivity,  Eq.~(\ref{sigmaLargeMu}), is valid for an arbitrary relation between the inelastic intra-valley scattering rate $1/\tau_\epsilon$ and the characteristic inter-valley scattering rate.
Since in the quasiclassical regime $\nu(\epsilon)\sim \epsilon^{2}$, the integral in  Eq.~(\ref{sigmaFormQuasi}) diverges at small $\epsilon$ and
should be cut off at energies of the order of electron level broadening $\gamma$. However, at $\mu \gg T$ the contribution to the conductivity from the small electron energies is exponentially small, $\sim e^{-\mu/T}$.

\emph{Nondegenerate regime.} At  $T\gg \mu$  inelastic scattering of electrons may no longer be ignored in the consideration of electrical conductivity.
In this regime the inelastic scattering rate is dominated by electron-electron collisions and may be estimated as
\begin{equation}
\frac{1}{\tau_{\epsilon}}\sim \alpha^{2}T.
\end{equation}
If the inelastic intra-valley scattering rate $1/\tau_{\epsilon}$ exceeds the characteristic rate of inter-valley scattering
$1/\tau(T)$  at $\epsilon\sim T$, then
the inelastic  scattering processes establish a locally equilibrium electron distribution function in a given valley
\begin{equation}\label{EqMu}
 n^{(a)}(\epsilon)=\frac{1}{\exp\left(\frac{\epsilon-\mu-\delta \mu^{(a)}}{T}\right)+1},
 \end{equation}
where the non-equilibrium part of the distribution is parameterized  by the correction $\delta \mu^{(a)}$ to the chemical potential. The value of $\delta \mu^{(a)}$  is controlled by the inter-valley relaxation time. One can find $\delta \mu^{(a)}$ with the aid of the electron number conservation law,
\begin{equation}\label{eq:NumberCons}
  \frac{\partial N^{(a)}}{\partial t}=k^{(a)}\frac{e^{2} ({\bf B\cdot E})}{4\pi^{2}\hbar^{2}c}+ \delta \mu^{(a)}
\int d\epsilon \frac{\nu^{(a)}(\epsilon)}{\tau(\epsilon)}\, \frac{\partial n_F(\epsilon)}{\partial \epsilon}
  =0,
\end{equation}
which follows from Eq.~(\ref{epsilonKinEq}).
Here
$
N^{(a)}=\int \nu^{(a)}(\epsilon)n^{(a)}(\epsilon) d \epsilon
$
 is the is the density of electrons in the $a$-th valley.
Substituting Eq.~(\ref{EqMu}) with  the obtained value of $\delta \mu$ into Eqs.~(\ref{epsilonCurrent}) and (\ref{IntegralCurrent}) we get the following  expression for the conductivity
\begin{equation}\label{sigmaT}
 \sigma_{zz}\sim \mathcal{N}_{v}\frac{e^{2}}{8\pi^{2}\hbar c}\frac{v}{c}\frac{( e  B)^{2}v^{2}}{T^{2}}\tau ({T}).
\end{equation}

Although this equation does not explicitly depend on the inelastic scattering rate $1/\tau_\epsilon$ it applies only if $\tau_\epsilon $ is sufficiently short.
Equation (\ref{sigmaT}) was obtained under the assumption that the nonequilibrium distribution is well approximated by the locally equilibrium form (\ref{EqMu}) with a valley-dependent chemical potential. This assumption breaks down at sufficiently low energies because the first term in the right hand side of Eq.~(\ref{epsilonKinEq}) grows as $1/\nu(\epsilon)$ when the energy decreases. Therefore the contribution to the conductivity from the small energy interval requires a special treatment. Using Eq.~(\ref{epsilonKinEq}) one can estimate the correction to the locally equilibrium distribution  (\ref{EqMu})  at energies  $\epsilon\sim \gamma\sim 1/\tau_{\epsilon}\sim \alpha^{2}T$ as
\begin{equation}
\delta n(\epsilon\sim \gamma)\sim
\frac{e^{2}}{4\pi^{2}\hbar^{2}c}(EB)
\frac{\tau_{\epsilon}}{T\nu(\gamma)}.
\end{equation}
Substituting this estimate into Eqs.~(\ref{epsilonCurrent}) and (\ref{IntegralCurrent}) we find that the corresponding correction to the conductivity is dominated by electrons
in the energy interval $\epsilon\sim \gamma$,  and is  smaller than Eq.~(\ref{sigmaT}) provided
\begin{equation}\label{ineq}
\frac{\tau_{\epsilon}}{\tau (T)}T\tau_{\epsilon} \sim \frac{\tau_{\epsilon}}{\tau (T)\alpha^{2}}\ll 1.
\end{equation}
Under the opposite condition, $\tau_\epsilon/(\tau (T) \alpha^2)\gg 1$,  as well as in the regime $\tau (T)<\tau_{\epsilon}$  the part of the conductivity related to the chiral anomaly
is determined by electrons in the energy interval $\epsilon\sim \gamma$ where
the electron level broadening becomes comparable to the energy, and the approach based  on quasiclassical Boltzmann equation is not applicable.
In the regime  where the level broadening is dominated by inelastic electron-electron collisions and satisfies $\gamma>v/L_{B}$ the conductivity may be estimated as
\begin{equation}\label{sigmaT1}
\sigma_{zz}\sim \mathcal{N}_{v}\frac{e^{2}}{8\pi^{2}\hbar c}\frac{v}{c}\frac{( e  B)^{2}v^{2}}{\hbar T}\tau_{\epsilon}^{2}.
\end{equation}

\emph{Einstein relation.}  Using  Eqs.~(\ref{epsilonCurrent}) and (\ref{epsilonKinEq}) one can also evaluate the electron diffusion coefficient. For example, at $T\ll \mu$,  assuming that $n^{(a)}({\bf r},\epsilon)=\bar{n}({\bf r},\epsilon)+\delta n^{(a)}({\bf r},\epsilon)$, ($\delta n^{(a)}({\bf r})\ll \bar{n}$, and $\sum_{a}\delta n^{(a)}({\bf r})=0$), at ${\bf E}=0$ we get
\begin{equation}\label{delta_n}
\delta n^{(a)}(\epsilon)= -k^{(a)}\frac{e}{4\pi^{2}\hbar^{2}c}\frac{\tau(\epsilon)}{\nu^{(a)}(\epsilon)}
 {\bf B} \cdot \frac{\partial \bar{n}({\bf r},\epsilon)}{\partial {\bf r}}
\end{equation}
Here $\bar{n}$ is the averaged over the valley part of the distribution function.
Substituting Eq.~(\ref{delta_n}) into Eq.~(\ref{epsilonCurrent}), and summing the result over $a$ we get
 the electron diffusion coefficient, which is  consistent with the Einstein relation.
\begin{equation}\label{DiffCoeff}
D^{(ch)}_{zz}(\mu)=\frac{\sigma_{zz}(\mu)}{e^{2}\nu(\mu)},
\end{equation}
where $\sigma_{zz}(\mu)$ is given by Eq.~(\ref{sigmaLargeMu}).

\emph{Thermal conductivity.} Similarly, at $\mu\gg T$, in the presence of a temperature gradient using Eq.~(\ref{epsilonCurrent}) we get
\begin{equation}\label{delta_n_kapa}
\delta n^{(a)} (\epsilon)= k^{(a)}\frac{\tau(\mu)}{\nu^{(a)}(\epsilon)} \frac{e}{4\pi^{2}\hbar^{2}c}\frac{\epsilon-\mu}{T}\,
\frac{\partial \bar{n}_{F}}{\partial \epsilon}{\bf B} \cdot \frac{\partial T}{\partial {\bf r}}.
\end{equation}
This yields the expression for the chiral-anomaly-related contribution to the longitudinal  thermal conductivity, which is consistent with the Wiedemann-Franz law,
\begin{equation}\label{kappaEl}
\kappa_{zz}= \frac{\pi^{2}\sigma_{zz}T}{3e^{2}}.
\end{equation}

At low magnetic fields in the quantum degenerate regime, $\mu \gg T$, the phonon contribution  to the thermal conductivity
is small compared to the electron one, as the phonon mean free path is limited by their absorption by electrons.
In the opposite limit $T,\mu\ll v/L_{B}$ of ultra-quantum magnetic field electrons can absorb phonons only in a narrow, of order $\gamma$,  interval of angles between electron and phonon momenta. In this case the thermal conductivity can be determined by the phonon transport.

\emph{Thermoelectric coefficient.} Using Eq.~(\ref{delta_n_kapa})  we can also get an expression for the thermoelectric coefficient which relates the electric current $j_{z}=\eta_{zz}(\nabla T)_{z}$ and the temperature gradient in z-direction. At $\mu \gg T$ this yields an expression consistent with the  Mott relation,
\begin{equation}
\eta_{zz}=\frac{\pi^{2}}{3e}T\frac{\partial \sigma_{zz}(T=0,\mu)}{\partial \mu}.
\end{equation}

\section{collective modes in Weyl semimetals\label{sec:collective}}

In this section we study manifestations of the chiral anomaly in the collective modes in Weyl semimetals.

\subsection{Magneto-plasmons and magneto-polaritons in Weyl semimetals\label{sec:plasmon}}

Knowledge of the frequency dependence of the conductivity $\sigma_{ij}(\omega)$ is sufficient to determine the plasmon spectrum.
In the quasiclassical limit where $max(T,\mu)\gg v/L_{B}$ and at $\omega\tau_{intra}\gg 1$ the conductivity is determined by all electrons, and  the contribution of the electrons in the chiral Landau level
is negligible. As a result the plasmon mode has a conventional for metals form.~\cite{Pesin}
In the ultra-quantum limit $\mu,T\ll v/L_{B}$ the situation is different.  Since at $\mu, T, B=0$  the plasmon mode does not exist, at $B\neq 0$  the plasmon spectrum is entirely determined by the chiral anomaly, and its form is very different from the  conventional plasmon spectrum.~\cite{SonSpivak}
Let us consider electromagnetic waves with frequencies below $v/L_{B}$. In this regime we may neglect excitations of electrons to higher Landau levels.
At $\omega\gg 1/\tau$  the ac conductivity is given by
\begin{equation}\label{eq:HighOmegaSigma}
 \sigma_{ij} = \frac{i \omega^2_0 }{4\pi \omega} n_i n_j,
\end{equation}
where $\n$ is the unit vector along the magnetic field and we  introduced the notation,
\begin{equation}\label{eq:omega_0}
  \omega_0^2=\mathcal{N}_v\frac{\pi e^2 v}{(\pi L_B)^2 \hbar }.
\end{equation}
Combining Eq.~(\ref{eq:HighOmegaSigma}) with the Poisson equation and the continuity equation we get the plasmon spectrum\cite{SonSpivak}
\begin{equation}\label{eq:plasmon_spectrum}
  \omega^2=\frac{\omega_0^2 q_z^2}{q_z^2+q_\perp^2}.
\end{equation}
Note that the plasmon frequency depends only on the angle between the wavevector $\q$ and the magnetic field, but not on its magnitude $|\q|$. The dependence of the plasmon frequency on the magnetic field is described by Eq.~(\ref{eq:omega_0}).

\begin{figure}
  \includegraphics[width=8cm]{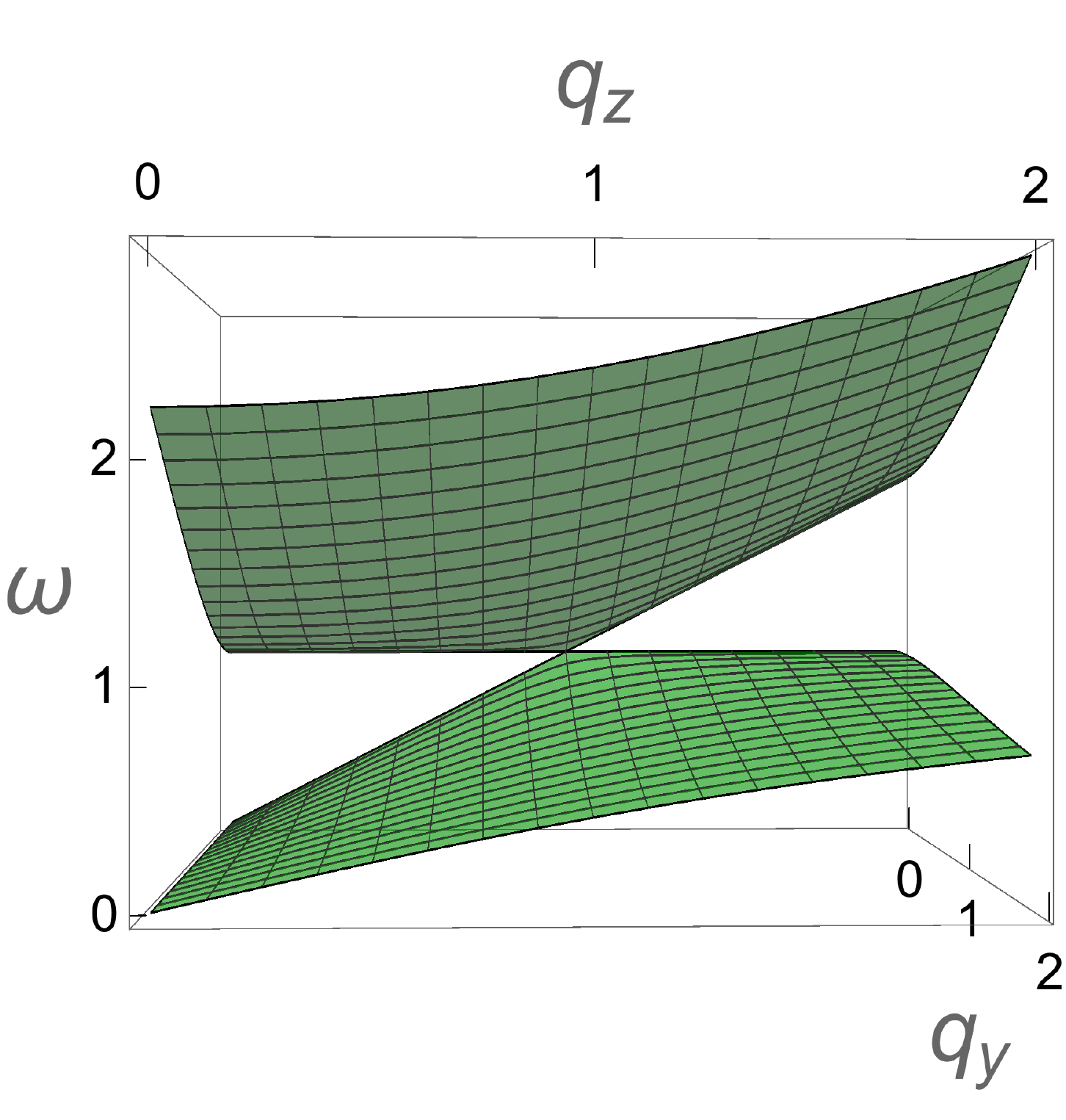}\\
  \caption{Spectrum of the two branches in the polariton region. The frequency $\omega$ (in units of $\omega_0$) is plotted as a function of the wave the wave vector (in units of $\omega_0/c$) at $q_x=0$.  The $q_z$-axis is directed from right to left, and the $q_y$-axis into the page. The two branches cross at $q_y=0$ and $q_z=\omega_0/c$. }\label{fig:polariton_spectrum}
\end{figure}

Equation~(\ref{eq:plasmon_spectrum}) was obtained in the approximation where the speed of light $c\rightarrow \infty$. At finite $c$ the hybridization  between the photon and the plasmon modes produces a polariton spectrum. The dispersion and polarization of electromagnetic waves, $\mathbf{E}(\mathbf{r},t)= \Re \mathbf{E}e^{- i \omega t + i \mathbf{q}\cdot \mathbf{r}}$ is determined by the equation~\cite{Landau8}
\begin{equation}\label{eq:transverse_E}
\left(  q_i q_j-q^2\delta_{ij}+ \frac{\omega^2}{c^2}\varepsilon_{ij}\right)E_j=0,
\end{equation}
where $\varepsilon_{ij}=\delta_{ij}+ 4\pi i \sigma_{ij}/\omega$ is the dielectric tensor. We denote the angle between the wave vector $\mathbf{q}$ and the $z$ axis by $\theta$, so that $\cos \theta = q_z/q$.
The wave polarized perpendicular to the $z$ axis does not produce and electric current and maintains the same dispersion as in vacuum, $\omega=c q$. The waves with the electric field polarized in the plane spanned by wavevector $\mathbf{q}$ and the $z$ axis represent superpositions of longitudinal (plasmon) and transverse (photon) waves.
For these waves  we obtain from Eq.~(\ref{eq:transverse_E})  the spectrum
\begin{equation}\label{eq:x_q}
  \omega^2=\frac{\omega^2_0+c^2 q^2}{2} \pm \frac{1}{2}\sqrt{(\omega^2_0 + c^2 q^2)^2 - 4 \omega_0^2 c^2 q^2 \cos^2 \theta}.
\end{equation}
which is plotted in Fig.~\ref{fig:polariton_spectrum}.
The plasmon spectrum, Eq.~(\ref{eq:plasmon_spectrum}), and unperturbed photon spectrum, $\omega=\pm cq$, are recovered from Eq.~(\ref{eq:x_q}) at $q\gg\omega_{0}/c$.  We note that the non-analytic in $\q$  dispersion of the plasmon in Eq.~(\ref{eq:plasmon_spectrum}) can be traced to the nonanalyticy of the spectrum (\ref{eq:x_q}) in the polariton region.

\subsection{Sound absorption in Weyl semimetals\label{sec:sound}}

As a sound wave propagates through a medium its intensity attenuates with the distance $r$ into the medium,  $I =I_0 e^{- \Gamma r}$, where $\Gamma$ is the absorption coefficient. In metals sound attenuation is dominated by the coupling  of the lattice deformation to the conduction electrons.  The strong anisotropic magnetic field dependence of the sound absorption coefficient $\Gamma$  also can reveal the significance of the chiral anomaly in Weyl semimetals.
For simplicity we consider a   sound wave
characterized by a lattice displacement vector $u_{i}({\bf r}, t)\sim \exp(i{\bf q}\cdot \r-i\omega t)$, where $\omega=c_{s}q$, and $c_{s}$ is the sound speed.
Sound absorption in Weyl semimetals is governed by the standard electron-lattice interaction, described by the deformation potential. We restrict ourselves to the relatively low frequency limit
$\omega \ll 1/\tau_{intra}, \omega_{p}$. In this case the charge oscillations induced by the strain are screened, while the corresponding oscillations of the scalar potential cancel the part of the electron energy proportional to the deformation potential
averaged over the valley of the electron spectrum (see for example Ref.~\onlinecite{Abricosov}).  As a result
the sound absorption is controlled by the effective deformation potential
\begin{equation}
\tilde{\Lambda}^{(a)}_{ij}=\bar{\Lambda}^{(a)}_{ij}-\frac{1}{\mathcal{N}_{v}}\sum_{i}\bar{\Lambda}^{(a)}_{ij}
\end{equation}
 (see for example Ref.~\onlinecite{Abricosov}).
According to Eq.~(\ref{epsilonKinEq}) $-\tilde{\Lambda}^{(a)}_{ij}\nabla u_{ij}({\bf r})/e$ acts as electric field acting on electrons in $a$-th valley. Let us introduce the electrical conductivity $\sigma^{(a)}_{zz}$ which can be interpreted as a conductivity associated with  valley $a$,  $\sigma_{zz}=\sum_{a}\sigma^{(a)}_{zz}$.
Then we can relate the entropy production to $\sigma^{(a)}_{zz}$
$T\dot{S}=\sum_{a}\sigma^{(a)}_{zz}\cos(\theta)q^{2}\langle (u_{ij} \tilde{\Lambda}^{(a)}_{ij})^{2}\rangle /e^{2}$.
   Dividing it by the sound wave energy density $\rho c_s^{2}\langle u_{ij}^{2}\rangle /2$, where   $\rho$ is the crystal density and $\langle \ldots \rangle$ denotes averaging over the period of oscillations,  at $\omega\tau\ll 1$ we get an expression for the sound absorption coefficient
\begin{equation}\label{soundAbSigma}
\Gamma= \frac{\sum_{i}\sigma^{(a)}_{zz}\cos^{2}(\theta)q^{2}(\tilde{\tilde{\Lambda}}^{(a)})^{2}/e^{2}}{e^{2}\rho c_s^{2}}\sim \sigma_{zz}\, \frac{\cos^{2}(\theta)q^{2}\tilde{\tilde{\Lambda}}^{2}/e^{2}}{e^{2}\rho c_s^{2}}
\end{equation}
where  $\theta$ is the angle between ${\bf B}$ and ${\bf q}$, $(\tilde{\tilde{\Lambda}}^{(a)})^{2}= \langle (u_{ij} \tilde{\Lambda}^{(a)}_{ij})^{2}\rangle/\langle u_{ij}^{2}\rangle$ and $\tilde{\tilde{\Lambda}}$ is a characteristic value of $\tilde{\tilde{\Lambda}}^{(a)}$.
Thus the sound absorption coefficient is strongly anisotropic function of the angle between ${\bf q}$ and ${\bf B}$.

\section{Discussion}

We have considered  the  electron transport phenomena in Weyl metals which are related to the  chiral anomaly.
 In our approximation the only nonvanishing components of the transport coefficient tensors are $\sigma_{zz}$, $\kappa_{zz}$ and $\eta_{zz}$.  Their strong  magnetic field dependence is controlled by the inter-valley scattering time $\tau$, and energy relaxation time $\tau_{\epsilon}$. Of course, there  are also conventional contributions to the transport coefficients
  which are related to the anisotropic part of the intra-valley distribution functions, and which are controlled by the intravalley relaxation time $\tau_{intra}$. The former contributions dominate the transport coefficients if the parameter $\tau/\tau_{intra}$ is sufficiently large.~\footnote{We note that our results are drastically different from those of Ref. \onlinecite{Fiete} which are proportional to the intra-valley relaxation time.}
For example, at low magnetic fields in  the quasiclassical regime, the chiral anomaly contribution to $\sigma_{zz}$ exceeds  the conventional Drude contribution if
\begin{equation}
\tau/\tau_{intra}>\tilde{N}_{B}\sim \left(\frac{\mathrm{max}(\mu,T)}{v/L_{B_{1}}}\right)^{2} \gg 1.
\end{equation}
 Here $\tilde{N}_{B}$ is the number of Landau levels in the energy interval $[0, \mathrm{max}(\mu,T)]$, and $B_{1}$ is the magnetic field where the resistance decreases by a factor of order one compared to it's zero magnetic field value.
 Even if chiral anomaly-related corrections are smaller than the Drude contribution to the conductivity, they still can dominate its magnetic field dependence provided
 \begin{equation}
 \frac{\tau(\mu)}{\tau_{intra}}\frac{1}{(\mu \tau_{intra})^{2}}\gg 1.
 \end{equation}

 The origin of the large parameter $\tau/\tau_{intra}$ in Weyl semimetals  may be related to the fact that the scattering potential is sufficiently smooth, and its inverse correlation radius is smaller than the value of the momentum transfer for the electron inter-valley scattering. In experiments \cite{Qiang Li_Kharzeev,JunXiong_Ong,Jun Xiong,Hasan,Xiaojun Yang,ChenZangEnzeZhangYanwenLiu} a strong linear in $B$ positive magnetoresistance was observed  at ${\bf E \perp B}$.
 A possible explanation of this phenomenon, associated with motion of electrons in quasiclassical magnetic field in the presence of smooth potential, was suggested in Ref.~\onlinecite{Lee}. This supports the picture that
 the  large parameter $\tau(\mu)/\tau_{intra}\gg 1$ originates from smooth disorder.
 We note however that the significant negative magnetoresistance for $\mathbf{E} \parallel \mathbf{B}$ has been also observed in Dirac semimetals, where the Dirac points in the electron spectrum are double degenerate. In this case the origin of the large parameter $\tau/\tau_{intra}$ is less clear.

In the framework of the conventional Boltzmann kinetic equation which does not take into consideration the existence of the Berry phase the magnetoresistance is always positive.~\cite{Abricosov,Lifshitz,Pippard} The negative magnetoresistance is small and isotropic in 3D systems at $p_{F}l\gg \hbar$, where $p_F$ is the Fermi momentum. We are not aware of other mechanisms of strong anisotropic negative magnetoresistance in conductors at low magnetic fields in the quasiclassical regime.
At high magnetic field, in the ultra-quantum limit, when only the zeroth Landau level is occupied, there is an unrelated to the chiral anomaly mechanism of strongly anisotropic magnetoresistance, which may become negative in the longitudinal direction  (see for example Refs.~\onlinecite{ArgyresAdams,Murzin,Sarma}).
As far as we know, this effect has never been observed in conventional semiconductors.
An additional difficulty in interpreting magnetotransport measurements in the ultra-quantum regime is associated with the instability of the electron liquid  with respect to charge density wave formation, which drives the system to the insulating state.
In contrast, in the semiclassical limit, theoretical consideration of electron transport is free of aforementioned complications.

In many experiments in Weyl semimetals for $\mathbf{E}\parallel \mathbf{B}$ (see for example Ref.~\onlinecite{Cai-ZhenLi}) the strong negative  magnetoresistance is preceded by a small positive magnetoresistance at relatively small fields $B< B^*$.
We believe, that the value of $v/L_{B^{*}}$ in these experiments is sufficiently low compared to $\mathrm{max}(T,\mu)$, so that the system may be treated quasiclassically in part of the region where negative magnetoresistance is observed. In addition we note that in some experiments (see for example Ref.~\onlinecite{Cai-ZhenLi}) there is an interval of temperatures where the negative magnetoresistance starts at zero magnetic field ($B^{*}=0)$, where the system, definitely can be described quasiclassically. Therefore we believe that the anisotropic magnetoresistance observed in the aforementioned experiments  is due to the chiral anomaly in these materials.

Another difference between the properties of magnetoresistance in quasiclassical and ultra-quantum regimes is that they have different dependence on the orientation of the magnetic field with respect to the crystalline axes.
At low magnetic field in
the quasiclassical regime the magnetoresistance related to the chiral anomaly is independent of the orientation of the magnetic field with respect to crystalline axis in spite  of the fact $\epsilon_{\bf p}$ and $\mathbf{\Omega}_{{\bf p}}$ are anisotropic functions of ${\bf p}$. The magnetic field dependence of the sound absorption coefficient exhibits similar properties. We would like to mention however
that the coefficient $\tilde{{\tilde \Lambda}}^{(a)}$ in Eq.~(\ref{soundAbSigma}) depends on the relative orientation of the vector ${\bf q}$ with respect to the crystalline axis.
In the the ultra-quantum case the value of the velocity in the direction of the magnetic field,
  the relaxation time in Eq.~(\ref{sigmaUltraQuant}), and consequently the conductivity $\sigma_{zz}$  depend on the magnetic field orientation.

\begin{acknowledgements}
We are grateful to D. T. Son and N. P. Ong for useful discussions.
The work of A. A. was supported by the US Department of Energy Office of Science, Basic Energy Sciences under award number
DE-FG02-07ER46452. B.S. thanks the International Institute of
Physics (Natal, Brazil) for hospitality.
\end{acknowledgements}

\end{document}